\documentclass[aps,pre,amsmath,amssymb,twocolumn]{revtex4}

\usepackage{graphicx}
\usepackage{amsmath}
\usepackage{verbatim}
\usepackage{xcolor} % for comments ...

\usepackage{algorithmic}
\usepackage{algorithm}

\newcommand{\iu}{\mathrm{i}} % imaginary unit
\DeclareMathOperator{\sech}{sech}
\DeclareMathOperator{\Op}{Op}

\begin{document}
\title{Reflectionless propagation of Manakov solitons on a line:\\ A model based on the concept of transparent boundary conditions}
\author{K.K. Sabirov$^{1,6}$, J.R. Yusupov$^{2}$, M.M. Aripov$^{3}$, M. Ehrhardt$^{4}$ and D.U. Matrasulov$^5$}
\affiliation{$^1$Tashkent University of Information Technologies, 108 Amir Temur Str., 100200, Tashkent Uzbekistan\\
$^2$Yeoju Technical Institute in Tashkent, 156 Usman Nasyr Str., 100121, Tashkent, Uzbekistan\\
$^3$National University of Uzbekistan, 2 Universitet Str., 100095, Tashkent, Uzbekistan\\
$^4$Bergische Universit\"at Wuppertal, Gau{\ss}strasse 20, D-42119 Wuppertal, Germany\\
$^5$Turin Polytechnic University in Tashkent, 17 Niyazov Str., 100095, Tashkent, Uzbekistan\\
$^6$Tashkent State Technical University, 2 Universitet Str., 100095, Tashkent, Uzbekistan}

%%%%%%%%%%%%%%%%%%%%%%%%%%%%%%%%%%%%%%
\begin{abstract}
We consider the problem of absence of backscattering in the transport of Manakov solitons on a line. The concept of transparent boundary conditions is used for modeling the reflectionless propagation of Manakov vector solitons in a one-dimensional domain. Artificial boundary conditions that ensure the absence of backscattering are derived and their numerical implementation is demonstrated. 

\end{abstract}

\maketitle

%%%%%%%%%%%%%%%%%%%%%%%%%%%%%%%%%%%%%%%%%%%%%
\section{Introduction}
The Manakov system is an integrable system of coupled nonlinear Schrödinger equations (NLSE) which allows diffe\-rent soliton solutions.  
The main application of the Ma\-na\-kov system comes from nonlinear optics, where it describes optical vector solitons propagating in Kerr media \cite{Kivshar1,Agarwal1}.
Manakov-type vector solitons also appear in optical fibers, in Bose-Einstein condensation, and in other areas of physics (see, e.g., Refs.~\cite{Kivshar1,Agarwal1} for an overview). 
So far, different aspects of the vector solitons described by the Manakov system have been studied~\cite{Lakshmanan1,Lakshmanan2,Ablowitz1,Lakshmanan3,Kivshar2,Lakshmanan4,Feng,Ohta,Panos1,Frantzeskakis}. 
In \cite{Kang} an experimental realization of such solitons in crystals was studied. 
The effect of small perturbations on the collision of vector solitons and its application to ultrafast soliton switching devices was investigated in \cite{Yang99}. 
The realization of logic gates and computational operations using Manakov vector solitons was discussed in \cite{Steiglitz}. 
The suppression of Manakov soliton interference in optical fibers caused by the interaction of two vector solitons in the Manakov equations that govern pulse transmission in randomly birefringent fibers was studied in \cite{Yang02}. Ref.~\cite{Zhong} studies the rogue waves described by the Manakov system with variable coefficients and external potential.  

In most cases, vector solitons are used as signal carriers in optics and optoelectronic technologies. 
For effective signal transmission in such devices and optimization of their functional properties, signal losses must be avoided or minimized by achieving a minimum of soliton backscattering, i.e., by propagating the solitons without reflections. 
The successful solution of such a problem requires the construction of mathematical models describing the reflectionless transport of solitons in a given medium. 
One of the effective mathematical tools for solving the problem of reflectionless soliton propagation is the imposition of so-called \textit{transparent boundary conditions (TBCs)} on a wave equation describing soliton transport. 
For special cases of the NLSE it is possible to formulate the exact TBC in closed form, cf.\ \cite{Zheng2006}.
For nonlinear wave equations, the concept of TBCs
%transparent boundary conditions 
often relies on the so-called \textit{unified approach} \cite{li2011} that is based on a splitting procedure of the linear and nonlinear part.
However, in general, for NLSE-type equations, it has turned out that the so-called \textit{potential approach} \cite{Antoine} is the most tractable one.
Here we extend this promising concept to the integrable Manakov system, which is a coupled system of NLSE with vector soliton solutions. 

This paper is organized as follows. 
In the next section we briefly recall soliton solutions and conserving quantities for the Manakov system on a line. 
Section~III presents the derivation of the transparent boundary conditions for the Manakov system. In Section~IV we demonstrate our numerical implementation of such complicated boundary conditions. Finally, Section~V provides the concluding remarks.

%%%%%%%%%%%%%%%%%%%%%%%%%%%%%%%%%%%%%%%%%%%%%
\section{Soliton solutions of the Manakov system}
The \textit{Manakov system} can be written explicitly as 
\begin{equation}\label{ms1}    
\begin{split}
\iu\partial_t\Psi_1+\frac{1}{2}\partial_x^2\Psi_1+(|\Psi_1|^2+|\Psi_2|^2)\Psi_1&=0,\\
\iu\partial_t\Psi_2+\frac{1}{2}\partial_x^2\Psi_2+(|\Psi_1|^2+|\Psi_2|^2)\Psi_2&=0,     
\end{split}
\end{equation}
where $(\Psi_1,\Psi_2)=\bigl(\Psi_1(x,t),\Psi_2(x,t)\bigr)$, $x\in\mathbb{R}$, $t>0$.
It was introduced first by Manakov \cite{Manakov} to describe statio\-nary self-focusing electromagnetic waves in homogeneous waveguide channels.  
The \textit{one-soliton solution} of the Manakov system can be written as 
\begin{equation}
  (\Psi_1^*, \Psi_2^*) = \iu\Bigl( \frac{c}{|c|}\Bigr)
  \frac{\eta \exp|2\iu(\eta^2-\xi^2)t -2\iu x\xi|}{\cosh[2\eta(x+x_0 +2\xi t)]},
\end{equation}
where the initial position of the soliton is given by
$x_0 =\ln (2\eta/|c|)/2\eta$ and
the unit vector $\bigl( c \equiv (c_{11},c_{21})]\bigr)$ determines the polarization of the soliton. 
The parameters $\xi$ and $\eta$ denote the speed and amplitude of the soliton, respectively.

Multi-soliton solutions of the Manakov system can be obtained using Hirota's bilinearization method \cite{Lakshmanan5,Lakshmanan6}. Eq.~\eqref{ms1} approves two conserving quantities, such as the norm determined as
\begin{equation*}
    N=\int_{-\infty}^{\infty} \bigl(|\Psi_1|^2 +|\Psi_2|^2\bigr)\,dx
\end{equation*}
and the energy, which is given by, cf.\ \cite{ismail20081}
\begin{equation}\label{eq:ener}
    E=\int\limits_{-\infty}^{+\infty}{\Biggl(\sum_{m=1}^{2}{\frac12\left|\frac{\partial\Psi_m}{\partial x}\right|^2}-\frac{1}{2}\sum_{m=1}^{2}{\left|\Psi_m\right|^4}-\left|\Psi_1\right|^2\left|\Psi_2\right|^2\Biggr)dx}.
\end{equation}

In the following we will use these quantities for confirming reflectionless transmission of Manakov solitons through a given boundary.

%%%%%%%%%%%%%%%%%%%%%%%%%%%%%%%%%%%%%%%%%%%%%%%%%%%%%%%%%%%%%%%%%%%%%%%%%
\section{Transparent boundary conditions for the Manakov system}

The reflection of nonlinear waves at the boundary of a given domain is a practically important problem, the solution of which requires the use of an explicit solution of a wave equation describing these waves. 
The mathematical description of the absence of reflection at the boundary is a rather complicated task, since unlike quantum mechanics, 
no S-matrix theory exists for nonlinear waves. 
One of the effective approaches to solve such a problem can be done in the framework of the concept of transparent boundary conditions. 
Such transparent boundary conditions for evolution equations can be constructed by coupling the solutions of the initial value boundary problems 
(IVBPs) in the interior and exterior domains
 \cite{Arnold1998,Ehrhardt1999,Ehrhardt2001,Ehrhardt2002,Arnold2003,Jiang2004,Antoine2008,Ehrhardt2008,Matthias2008,Sumichrast2009,Antoine2009,Ehrhardt2010,Klein2011,Arnold2012}.

To construct transparent boundary conditions (TBCs) for a given PDE, one must first split the original wave equation into coupled equations determined in the interior ($\Omega^{\rm int}$) and exterior ($\Omega^{\rm ext}$) domains. 
Then one applies a Laplace transform in time to the exterior problems and obtains the solution of the ordinary differential equations in the spatial variable $x$.  
Moreover, if one allows only "outgoing" waves by choosing the asymptotically decaying solution as 
$x\to\pm\infty$ and matching the Dirichlet and Neumann values on the artificial boundaries of the interior domain, one should apply (numerically) the inverse Laplace transform to complete the full derivation of the TBC \cite{Antoine2008}.

Here we will apply the above concept and procedure for the derivation of TBCs for the Manakov system Eq.~\eqref{ms1} and their numerical implementation at the artificial boundary points $x=0$, $x=L$. 
For this purpose, we use the so-called \textit{potential approach}, which was previously used to derive TBCs for the nonlinear Schr\"odinger equation~\cite{Antoine,Jambul1} and the sine-Gordon equation \cite{TBCSGE}. 
In \cite{Jambul,Jambul02,Jambul2} the TBC concept was used  to develop transparent quantum graphs model, which was later implemented to describe reflectionless transport of charge carriers in branched conducting polymers~\cite{Exciton}.  
Within such an approach, the Manakov system~\eqref{ms1} is formally reduced to a system of linear PDEs by introducing the following \textit{potential}
 $V(x,t)=|\Psi_1|^2+|\Psi_2|^2$. 
 This procedure allows us to rewrite Eq.~\eqref{ms1} into the following linear form:
\begin{equation}\label{ms2}    
    \begin{split}
\iu\partial_t\Psi_1+\frac{1}{2}\partial_x^2\Psi_1+V(x,t)\Psi_1&=0,\\
\iu\partial_t\Psi_2+\frac{1}{2}\partial_x^2\Psi_2+V(x,t)\Psi_2&=0.
\end{split}
\end{equation}
%Furthermore, we introduce the following new vector function as
We also introduce the following new vector function as
\begin{equation} \label{nvf1}
    \begin{split}
  v(x,t)&=e^{-\iu\nu(x,t)}\Psi(x,t),\\
  \Psi(x,t)&=\begin{pmatrix}\Psi_1(x,t)\\\Psi_2(x,t)\end{pmatrix},\quad
  v(x,t)=\begin{pmatrix}v_1(x,t)\\v_2(x,t)\end{pmatrix},
\end{split}
\end{equation}
where
\begin{equation}  \label{eq:nu}
    \begin{split}
    \nu(x,t)&=\int_0^tV(x,\tau)\,d\tau\\
    &=\int_0^t\bigl(|\Psi_1(x,\tau)|^2+|\Psi_2(x,\tau)|^2\bigr)\,d\tau.
\end{split}
\end{equation}
Taking here partial derivatives we obtain
\begin{align*}
   \partial_t\Psi &=e^{\iu\nu}(\partial_t+\iu V)v,\\
   \partial_x^2\Psi &=e^{\iu\nu}\bigl(\partial_x^2+2\iu\partial_x\nu\cdot\partial_x+\iu\partial_x^2\nu-(\partial_x\nu)^2\bigr)v.
\end{align*}
Inserting these latter equations in \eqref{ms2},  we get
\begin{equation}
L(x,t,\partial_x,\partial_t)v=\iu\partial_tv
+\frac{1}{2}\partial_x^2v+A\partial_xv+Bv=0,\label{ms3}
\end{equation}
where
$A=\iu\partial_x\nu$, $B=\frac{1}{2}\bigl(\iu\partial_x^2\nu-(\partial_x\nu)^2\bigr)$.
Linearizing Eq.~\eqref{ms3} using pseudo-differential operators we have
\begin{multline}\label{ms4}
L=\Bigl(\frac{1}{\sqrt{2}}\partial_x+\iu\Lambda^-\Bigr)\Bigl(\frac{1}{\sqrt{2}}\partial_x+i\Lambda^+\Bigr)\\
=\frac{1}{2}\partial_x^2+\frac{\iu}{\sqrt{2}}(\Lambda^++\Lambda^-)\partial_x
+\frac{\iu}{\sqrt{2}}\Op(\partial_x\lambda^+)-\Lambda^-\Lambda^+.
\end{multline}
From Eqs.~\eqref{ms3} and \eqref{ms4} we obtain the following
system of operators
\begin{align}
\frac{\iu}{\sqrt{2}}(\Lambda^++\Lambda^-)&=A,\nonumber\\
\frac{\iu}{\sqrt{2}}\Op(\partial_x\lambda^+)-\Lambda^-\Lambda^+&=\iu\partial_t+B,\label{soop1}
\end{align}
which yields the symbolic system of equations
\begin{align}
  \frac{\iu}{\sqrt{2}}(\lambda^++\lambda^-)&=a,\nonumber\\
  \frac{\iu}{\sqrt{2}}\partial_x\lambda^+-\sum_{\alpha=0}^{+\infty}\frac{(-\iu)^\alpha}{\alpha!}
  \partial_\tau^\alpha\lambda^-\partial_t^\alpha\lambda^+&=-\tau+b,\label{sos1}
\end{align}
where the setting $a=A$, $b=B$ is used. An asymptotic expansion in the inhomogeneous
symbols is defined as
\begin{equation}
    \lambda^\pm\sim\sum_{j=0}^{+\infty}\lambda_{1/2-j/2}^\pm.\label{expl}
\end{equation}
Inserting the expansion \eqref{expl} into Eq.~\eqref{sos1} we can identify the terms of order 1/2 in the first relation of the system
\eqref{sos1}:
\begin{equation}\label{o12}
   \lambda_{1/2}^-=-\lambda_{1/2}^+,\quad
   \lambda_{1/2}^+=\pm\sqrt{-\tau}.
\end{equation}
The Dirichlet-to-Neumann  
operator corresponds to the choice
$\lambda_{1/2}^+=-\sqrt{-\tau}$. 
For the zero order terms we obtain
\begin{eqnarray}
\lambda_0^-=-\lambda_0^+-\iu\sqrt{2}a,\nonumber\\
\frac{\iu}{\sqrt{2}}\partial_x\lambda_{1/2}^+-(\lambda_0^-\lambda_{1/2}^++\lambda_0^+\lambda_{1/2}^-)=0.\label{zo1}
\end{eqnarray}
From Eq.~\eqref{zo1} we get
\begin{equation}\label{zo2}
\begin{split}
\lambda_0^+&=-\iu\frac{\sqrt{2}}{2}a=\frac{\sqrt{2}}{2}\partial_x\nu,\\
\lambda_0^-&=-\lambda_0^+-\iu\sqrt{2}a=\frac{\sqrt{2}}{2}\partial_x\nu.
\end{split}
\end{equation}
For the terms of order -1/2 we have
\begin{eqnarray}
\frac{\iu}{\sqrt{2}}(\lambda_{-1/2}^++\lambda_{-1/2}^-)=0,\nonumber\\
\frac{\iu}{\sqrt{2}}\partial_x\lambda_0^+-(\lambda_{1/2}^-\lambda_{-1/2}^++\lambda_0^-\lambda_0^++\lambda_{-1/2}^-\lambda_{1/2}^+)=b,\label{om12}
\end{eqnarray}
since
$\partial_t^\alpha\lambda_{-1/2}^\pm=\partial_\tau^\alpha\lambda_0^\pm=0,\,\alpha\in
N$. From \eqref{om12} we get
\begin{equation}
\lambda_{-1/2}^\pm=0.\label{om121}
\end{equation}
Furthermore, one can obtain the next order terms as
\begin{eqnarray}
\lambda_{-1}^-=-\lambda_{-1}^+,\,\lambda_{-1}^+=\iu\frac{\sqrt{2}}{8\tau}\partial_xV.\label{om1}
\end{eqnarray}
Then the first order approximation reads
\begin{eqnarray}
\frac{1}{\sqrt{2}}\partial_x\Psi_1|_{x=0}-e^{-\iu\frac{\pi}{4}}e^{\iu\nu}\cdot\partial_t^{1/2}(e^{-\iu\nu}\Psi_1)|_{x=0}=0,\nonumber\\
\frac{1}{\sqrt{2}}\partial_x\Psi_2|_{x=0}-e^{-\iu\frac{\pi}{4}}e^{\iu\nu}\cdot\partial_t^{1/2}(e^{-\iu\nu}\Psi_2)|_{x=0}=0,\label{foal}\\
\frac{1}{\sqrt{2}}\partial_x\Psi_1|_{x=L}+e^{-\iu\frac{\pi}{4}}e^{\iu\nu}\cdot\partial_t^{1/2}(e^{-\iu\nu}\Psi_1)|_{x=L}=0,\nonumber\\
\frac{1}{\sqrt{2}}\partial_x\Psi_2|_{x=L}+e^{-\iu\frac{\pi}{4}}e^{\iu\nu}\cdot\partial_t^{1/2}(e^{-\iu\nu}\Psi_2)|_{x=L}=0.\label{foar}
\end{eqnarray}
The second order approximation is
\begin{multline*}
\frac{1}{\sqrt{2}}\partial_x\Psi_1|_{x=0}-e^{-\iu\frac{\pi}{4}}e^{\iu\nu}\cdot\partial_t^{1/2}(e^{-\iu\nu}\Psi_1)|_{x=0}\\
-\iu\frac{\sqrt{2}}{8}\partial_xVe^{\iu\nu}I_t(e^{-\iu\nu}\Psi_1)|_{x=0}=0,
\end{multline*}
\begin{multline}\label{soal}
\frac{1}{\sqrt{2}}\partial_x\Psi_2|_{x=0}-e^{-\iu\frac{\pi}{4}}e^{\iu\nu}\cdot\partial_t^{1/2}(e^{-\iu\nu}\Psi_2)|_{x=0}\\
-\iu\frac{\sqrt{2}}{8}\partial_xVe^{\iu\nu}I_t(e^{-\iu\nu}\Psi_2)|_{x=0}=0,
\end{multline}
\begin{multline*}
 \frac{1}{\sqrt{2}}\partial_x\Psi_1|_{x=L}+e^{-\iu\frac{\pi}{4}}e^{\iu\nu}\cdot\partial_t^{1/2}(e^{-\iu\nu}\Psi_1)|_{x=L}\\
 +\iu\frac{\sqrt{2}}{8}\partial_xVe^{\iu\nu}I_t(e^{-\iu\nu}\Psi_1)|_{x=L}=0,
\end{multline*}
\begin{multline}\label{soar}
\frac{1}{\sqrt{2}}\partial_x\Psi_2|_{x=L}+e^{-\iu\frac{\pi}{4}}e^{\iu\nu}\cdot\partial_t^{1/2}(e^{-\iu\nu}\Psi_2)|_{x=L}\\
+\iu\frac{\sqrt{2}}{8}\partial_xVe^{\iu\nu}I_t(e^{-\iu\nu}\Psi_2)|_{x=L}=0,
\end{multline}
where $I_t(f)=\int_0^t{f(\tau)\,d\tau}$.
Unlike the standard the standard Dirichlet, Neumann or Robin boundary conditions, the boundary conditions given by Eqs.~\eqref{foar} and \eqref{soar}, are quite complicated and can be implemented only numerically. 
Therefore, we will provide in the next section their numerical implementation for the Manakov system \eqref{ms1}.

%%%%%%%%%%%%%%%%%%%%%%%%%%%%%%%%%%%%%%%%%%%
\section{Discretization of the Manakov system and transparent boundary conditions}

For the numerical solution of the system~\eqref{ms1} by imposing transparent boundary conditions given by Eqs.~\eqref{foar} and \eqref{soar} one must use an effective discretization scheme for both, Eq.~\eqref{ms1} and the  boundary conditions. 
In the case of transparent boundary conditions, the accuracy and stability of the numerical solution is very sensitive to the choice of a discretization scheme.  
Here we present a numerical scheme for Eq.~\eqref{ms1} and a procedure for implementing the transparent boundary conditions.

\subsection{Discretization of the equation}
The numerical solution of coupled Schr\"odinger equations is a well-studied problem and different high accuracy numerical methods have been developed in the literature so far (see, e.g.\ \cite{ismail2001,ismail2004,ismail2007,ismail2008,xu2010}).

Here we use the explicit mid point rule \cite{frutos1992}, the so-called
\textit{leap-frog finite difference method} which is given as
\begin{equation}\label{fds}
\begin{split}
\iu\frac{\Psi^{n+1}_{1,j}-\Psi^{n-1}_{1,j}}{2\Delta t}+\frac12 D_x^2 \Psi^{n}_{1,j}+V_j^n\Psi^{n}_{1,j}=0,\\
\iu\frac{\Psi^{n+1}_{2,j}-\Psi^{n-1}_{2,j}}{2\Delta t}+\frac12 D_x^2 \Psi^{n}_{2,j}+V_j^n\Psi^{n}_{2,j}=0,
\end{split}
\end{equation}
with the standard second order difference quotient
\begin{equation*}
    D_x^2\Psi_j^n=\dfrac{1}{\Delta x^2}(\Psi^n_{j+1}-2\Psi^n_{j}+\Psi^n_{j-1}),
\end{equation*}
and the discrete potential term
\begin{equation*}
    V_j^n=|\Psi^n_{1,j}|^2+|\Psi^n_{2,j}|^2,
\end{equation*}
where $\Delta t$ and $\Delta x$ are the time and space discretization steps, respectively. 
We note that there are other implicit or semi-implicit methods with higher accuracy~\cite{ismail2008,xu2010,frutos1992,zhou2010,chen2017,hu2020}. 
But the complexity of TBC forces to choose between accuracy and computational cost. 
And this method was chosen because of its simple implementation and low cost per step. 

This leads to the following explicit finite difference scheme: 
\begin{equation}\label{fds1}
    \mathbf{U}_j^{n+1}=\mathbf{U}^{n-1}_{j} + \iu \Delta t D_x^2\mathbf{U}^{n}_{j} + 2\iu\Delta t V_j^n\mathbf{U}^{n}_{j},
\end{equation}
where
\begin{equation*}
    \mathbf{U}^n_j=\begin{pmatrix}\Psi^n_{1,j} \\
                     \Psi^n_{2,j}\end{pmatrix}.    
\end{equation*}

Furthermore, we have to implement the transparent boundary conditions given by Eqs.~\eqref{foal}-\eqref{foar} or Eqs.~\eqref{soal}-\eqref{soar} in the above numerical methods.

%%%%%%%%%%%%%%%%%%%%%%%%%%%%%%%%%%%%%%%%%%%%%%%%%%%%%%%%%%%%%%%%%%%%%%%%%%%%%
\subsection{Implementation of the TBC}

The discretization of the TBC given by the Eqs.~\eqref{foal}-\eqref{foar} and its subsequent implementation in the nume\-rical scheme for the Eq.~\eqref{ms1} and ensuring the stability of the whole numerical scheme is a rather complicated task. The presence of the fractional derivative also makes the discretization scheme very complicated and unstable. 
Here we give an effective discretization scheme for the TBC, which can be implemented with high accuracy and stability in combination with the discretization scheme for Eq.~\eqref{ms1}. 
We state the scheme only for $x=L$ by saying that for the left boundary (at $x=0$) the implementation can be done in the same way.

The approximation of the fractional differential operator is given by the numerical quadrature formula \cite{Antoine}
\begin{equation*}
   \partial_t^{1/2}f(t_n)\approx\frac{2}{\sqrt{2\Delta t}}\sum\limits_{k=0}^n{\beta_k f^{n-k}},    
\end{equation*}
where $\{f_n\}_{n\in\mathbb{N}}$ is a sequence of complex values appro\-ximating $\{f(t_n)\}_{n\in\mathbb{N}}$ and $(\beta_k)_{k\in\mathbb{N}}$ denotes the sequences deﬁned by
\begin{multline*}
(\beta_0,\beta_1,\beta_2,\beta_3,\beta_4,\beta_5,\ldots)=\\
\left(1,-1,\frac12,-\frac12,\frac{1\cdot3}{2\cdot4},-\frac{1\cdot3}{2\cdot4},\ldots\right).
\end{multline*}
The function $\nu(x,t)$ given by \eqref{eq:nu} can be discretized using the trapezoidal rule as
\begin{equation*}
\nu^n_j=\Delta t\left[\sum\limits_{k=1}^{n-1}V_j^k+\frac12V_j^n\right], 
\quad\text{for}\quad n\ge2,
\end{equation*}
with $\nu^0_j=0$ and $\nu^1_j=\frac{\Delta t}{2}V_j^1$.
Let us note, that the term $\frac{1}{2} V_0^n$ was dropped, 
because the initial data is assumed to be compactly supported and hence $V^n_0$ is zero. 
Then, we discretize the function $e^{\iu\nu(x,t)}$ from \eqref{nvf1} as
\begin{multline*}
     E_j^n=\exp{(\iu\nu_j^n)}=\\
     \exp{\Bigl(\iu\Delta t\sum\limits_{k=1}^{n-1}V_j^k\Bigr)}\cdot\exp{\Bigl(\iu\Delta t\frac12V_j^n\Bigr)}\\
=\widetilde{E_j^{n-1}}\cdot\exp{\Bigl(\iu\Delta t\frac{1}{2}V_j^n\Bigr)},
\end{multline*}
where $\widetilde{E_j^{n-1}}=\exp{\Bigl(\iu\Delta t\sum\limits_{k=1}^{n-1}V_j^k}\Bigr)$.

Thus, the TBC operator of the first order approximation \eqref{foar}
at the right boundary $j=J$ can be approximated by the discrete convolutions
\begin{align*}
(\Lambda_1^n)_I&=e^{-\iu\frac{\pi}{4}}\sqrt{\frac{2}{\Delta t}}E_J^n\sum\limits_{k=0}^{n}{\beta_{k}\overline{E_J^{n-k}}\Psi_{1,J}^{n-k}},\\    
(\Lambda_2^n)_I&=e^{-\iu\frac{\pi}{4}}\sqrt{\frac{2}{\Delta t}}E_J^n\sum\limits_{k=0}^{n}{\beta_{k}\overline{E_J^{n-k}}\Psi_{2,J}^{n-k}},    
\end{align*}
where $\overline{E_J^n}$ denotes the complex conjugate of $E_J^n$.

Then the values of the wave function at the right boundary can be obtained by solving the system of nonlinear equations with respect to $(\Psi_{1,J}^n,\Psi_{2,J}^n)^\top$, given as
\begin{multline*}
\frac{\Psi_{1,J}^n-\Psi_{1,J-1}^n}{\Delta x}+e^{-\iu\frac{\pi}{4}}\frac{2}{\sqrt{\Delta t}}\Bigl[\Psi_{1,J}^n+\widetilde{E_J^{n-1}}\cdot\\\exp{\Bigl(\iu\frac{\Delta t}{2}\bigl(|\Psi_{1,J}^n|^2+|\Psi_{2,J}^n|^2\bigr)\Bigr)}\sum\limits_{k=1}^{n}{\beta_{k}\overline{E_J^{n-k}}\Psi_{1,J}^{n-k}}\Bigr]=0,
\end{multline*}
\begin{multline*}
\frac{\Psi_{2,J}^n-\Psi_{2,J-1}^n}{\Delta x}+e^{-\iu\frac{\pi}{4}}\frac{2}{\sqrt{\Delta t}}\Bigl[\Psi_{2,J}^n+\widetilde{E_J^{n-1}}\cdot\\\exp{\Bigl(\iu\frac{\Delta t}{2}\bigl(|\Psi_{1,J}^n|^2+|\Psi_{2,J}^n|^2\bigr)\Bigr)}\sum\limits_{k=1}^{n}{\beta_{k}\overline{E_J^{n-k}}\Psi_{2,J}^{n-k}}\Bigr]=0.
\end{multline*}

Using the same approach, we can proceed with the discretization of the second order approximation. 
We recall that $V(x,t)=\overline{\Psi}_1\Psi_1+\overline{\Psi}_2\Psi_2$ and  
approximate $\partial_x V(x,t)$ at the right boundary $x=L$ (i.e.\ $j=J$) with
\begin{align*}
dV_J^n=\frac{1}{\Delta x} \Bigl(&2|\Psi_{1,J}^n|^2-\overline{\Psi_{1,J-1}^n}\Psi_{1,J}^n-\Psi_{1,J-1}^n\overline{\Psi_{1,J}^n}+\\
&2|\Psi_{2,J}^n|^2-\overline{\Psi_{2,J-1}^n}\Psi_{2,J}^n-\Psi_{2,J-1}^n\overline{\Psi_{2,J}^n}\Bigr),
\end{align*}
where $\overline{\Psi}$ is the complex conjugate of $\Psi$. 
Then, again using the trapezoidal method, we approximate the integral term $I_t(\cdot)$ in \eqref{soar} %at $x=L$ 
with
\begin{equation*}
I_{m,t}^n=\Delta t \left(\sum\limits_{k=1}^{n-1}{\overline{E_J^k}\Psi_{m,J}^k}+\frac12\overline{E_J^n}\Psi_{m,J}^n\right),\ m=1,2.
\end{equation*}
Thus, the TBC operator of the second order approximation \eqref{soar} can be approximated as
\begin{align*}
(\Lambda_1^n)_{II}&=(\Lambda_1^n)_I+\iu\frac{\sqrt{2}}{8}dV_J^nE_J^nI_{1,t}^n,\\    
(\Lambda_2^n)_{II}&=(\Lambda_2^n)_I+\iu\frac{\sqrt{2}}{8}dV_J^nE_J^nI_{2,t}^n.
\end{align*}

Again, the values of the wave function at the right boundary can be obtained by solving the system of nonlinear equations with respect to $(\Psi_{1,J}^n,\Psi_{2,J}^n)^\top$, given as
\begin{multline*}
\frac{\Psi_{1,J}^n-\Psi_{1,J-1}^n}{\Delta x}+e^{-\iu\frac{\pi}{4}}\frac{2}{\sqrt{\Delta t}}\Bigl[\Psi_{1,J}^n+\widetilde{E_J^{n-1}}\cdot\\
\exp{\Bigl(\iu\frac{\Delta t}{2}\bigl(|\Psi_{1,J}^n|^2+|\Psi_{2,J}^n|^2\bigr)\Bigr)}\sum\limits_{k=1}^{n}{\beta_{k}\overline{E_J^{n-k}}\Psi_{1,J}^{n-k}}\Bigr]+\\
\iu\frac{\Delta t}{4\Delta x}\Bigl(2|\Psi_{1,J}^n|^2-\overline{\Psi_{1,J-1}^n}\Psi_{1,J}^n-\Psi_{1,J-1}^n\overline{\Psi_{1,J}^n}+\\
2|\Psi_{2,J}^n|^2-\overline{\Psi_{2,J-1}^n}\Psi_{2,J}^n-\Psi_{2,J-1}^n\overline{\Psi_{2,J}^n}\Bigr)\cdot\\\widetilde{E_J^{n-1}}\cdot\exp{\Bigl(\iu\frac{\Delta t}{2}\bigl(|\Psi_{1,J}^n|^2+|\Psi_{2,J}^n|^2\bigr)\Bigr)}\cdot\\\left(\sum\limits_{k=1}^{n-1}{\overline{E_J^k}\Psi_{1,J}^k}+\frac12\overline{E_J^n}\Psi_{1,J}^n\right)=0,
\end{multline*}
\begin{multline*}
\frac{\Psi_{2,J}^n-\Psi_{2,J-1}^n}{\Delta x}+e^{-\iu\frac{\pi}{4}}\frac{2}{\sqrt{\Delta t}}\Bigl[\Psi_{2,J}^n+\widetilde{E_J^{n-1}}\cdot\\\exp{\Bigl(\iu\frac{\Delta t}{2}\bigl(|\Psi_{1,J}^n|^2+|\Psi_{2,J}^n|^2\bigr)\Bigr)}\sum\limits_{k=1}^{n}{\beta_{k}\overline{E_J^{n-k}}\Psi_{2,J}^{n-k}}\Bigr]+\\\iu\frac{\Delta t}{4\Delta x}\Bigl(2|\Psi_{1,J}^n|^2-\overline{\Psi_{1,J-1}^n}\Psi_{1,J}^n-\Psi_{1,J-1}^n\overline{\Psi_{1,J}^n}+\\
2|\Psi_{2,J}^n|^2-\overline{\Psi_{2,J-1}^n}\Psi_{2,J}^n-\Psi_{2,J-1}^n\overline{\Psi_{2,J}^n}\Bigr)\cdot\\\widetilde{E_J^{n-1}}\cdot\exp{\Bigl(\iu\frac{\Delta t}{2}\bigl(|\Psi_{1,J}^n|^2+|\Psi_{2,J}^n|^2\bigr)\Bigr)}\cdot\\\left(\sum\limits_{k=1}^{n-1}{\overline{E_J^k}\Psi_{2,J}^k}+\frac12\overline{E_J^n}\Psi_{2,J}^n\right)=0.
\end{multline*}

\begin{figure}[th!]
\includegraphics[width=8.6cm]{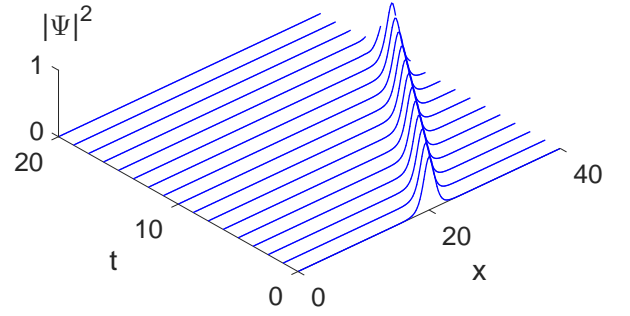}
\caption{Evolution of a right-travelling single soliton simulated with the finite difference scheme~\eqref{fds1}. 
The first order approximation of the TBC is imposed at the right boundary ($x=40$).}
\label{sd}
\end{figure}
    
\begin{figure}[th!]
\includegraphics[width=8.6cm]{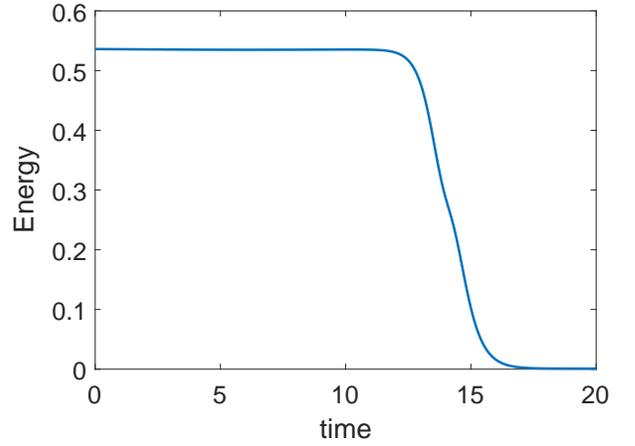}
\caption{Time evolution of the soliton energy in the interior (computational) domain $[0,L]$.}\label{ener}
\end{figure}

%%%%%%%%%%%%%%%%%%%%%%%%%%%%%%%%%%%%%%%%%%%
\section{Numerical Experiment}
We solve the system of coupled nonlinear Schr\"odinger equations, 
given by Eq.~\eqref{ms1} on the finite interval $[0, L]$ and impose the TBC at the right boundary (at $x=L$).
As initial condition we choose a  single soliton from the exact solution given by
\begin{align*}
&\Psi_1(x,0)=\sqrt{\alpha}\sech\bigl[\sqrt{2\alpha}(x-x_0)\bigr]\exp\bigl[\iu\sqrt{2}p(x-x_0)\bigr]\\     
&\Psi_2(x,0)=\sqrt{\alpha}\sech\bigl[\sqrt{2\alpha}(x-x_0)\bigr]\exp\bigl[\iu\sqrt{2}p(x-x_0)\bigr].
\end{align*}

In our experiments we selected the following system parameters: $L=40$, the parameters of the initial condition 
$\alpha=1$, $p=1$, $x_0=20$ and the discretization parameters $\Delta x=0.05$, $\Delta t=0.00125$. 
The evolution of the right-travelling single soliton is shown in Fig.~\ref{sd}.

To check the the absence of back scattering, we numerically calculate and plot the time dependence of the energy given by Eq.~\eqref{eq:ener}. 
The fact that the energy becomes zero while time elapses can be considered as a marker of the TBC. Assuming that the wave function determined by the initial conditions is negligibly small outside the computational domain $[0, L]$, Eq.~\eqref{eq:ener} can be rewritten as
\begin{equation}\label{eq:ener1}
    E=\frac12\int\limits_{0}^{L}{\left(\sum\limits_{m=1}^{2}{\left|\frac{\partial\Psi_m}{\partial x}\right|^2}-\left(\left|\Psi_1\right|^2+\left|\Psi_2\right|^2\right)^2\right)dx}.
\end{equation}

In our calculations we use the discrete analog of the energy ($E(t_n)=E_n$) given by Eq.~\eqref{eq:ener1}:
\begin{align*}
E_n=\frac{1}{4\Delta x}\sum\limits_{j=1}^{J-1}&\Bigl[\left|\Psi_{1,j+1}^n-\Psi_{1,j-1}^n\right|^2+\left|\Psi_{2,j+1}^n-\Psi_{2,j-1}^n\right|^2\\
&-2\Delta x^2\left(\left|\Psi_{1,j}^n\right|^2+\left|\Psi_{2,j}^n\right|^2\right)^2\Bigr].
\end{align*}

The time evolution of the soliton energy within the limits of the calculation interval is shown in Fig.~\ref{ener}. 
It is clear that the energy disappears when the soliton crosses the boundary, i.e.\ the absence of backscattering. 

\begin{figure}[th!]
\includegraphics[width=8.6cm]{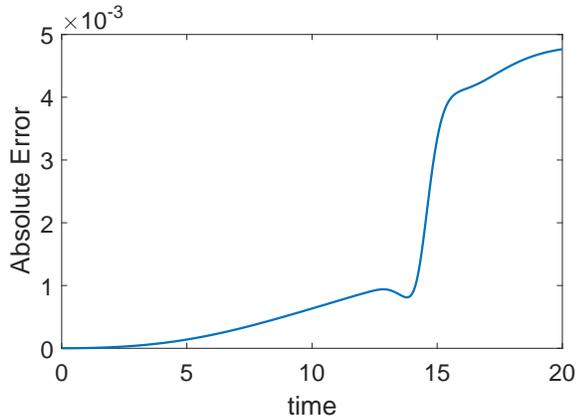}
% \caption{Plot of the absolute error measured by the $L^2$-norm.}
\caption{Plot of the absolute error 'Err' in the $L^2$-norm.}\label{err}
\end{figure}

At the end, to show that the numerical solution is reliable, we calculate the absolute error 'Err' measured with the $L^2$-norm (discretized by the  trapezoidal rule):
\begin{multline*}
||\text{Err}(t_n)||_2^2=\Delta x \sum\limits_{j=1}^{J-1}\Bigl(\bigl|\Delta\Psi_{1,j}^n\bigr|^2+\bigl|\Delta\Psi_{2,j}^n\bigr|^2\Bigr)+\\ \frac{\Delta x}{2}\Bigl(\bigl|\Delta\Psi_{1,0}^n\bigr|^2+\bigl|\Delta\Psi_{2,0}^n\bigr|^2+\bigl|\Delta\Psi_{1,J}^n\bigr|^2+\bigl|\Delta\Psi_{2,J}^n\bigr|^2\Bigr),
\end{multline*}
where $\Delta\Psi_{k,j}^n=\Psi_{k,j}^n-\Psi_k^{\text{exact}}(x_j,t_n)$. The plot of this error over time is presented in Fig.~\ref{err}.

%%%%%%%%%%%%%%%%%%%%%%%%%%%%%%%%%%%%%%%%%%%%%
\section{Conclusions}
In this work, the reflectionless transmission of Manakov solitons described by the Manakov system on a line 
subject to so-called transparent boundary conditions has been studied. 
These transparent boundary conditions (TBCs) are derived analytically and an effective discretization for the TBCs and its implementation in the numerical method for the Manakov system are presented. 
The absence of backscattering for Manakov solitons, when TBCs are imposed is demonstrated by a direct numerical experiment. The results of the work can be used for modeling the tunable transport of Manakov solitons in optical media and for optimization problems of optical devices, where such solitons appear as signal carriers. The above model can be extended for modeling the reflectionless propagation of Manakov vector solitons  in higher dimensional and branched domains, which are now the tasks under process.

%%%%%%%%%%%%%%%%%%%%%%%%%%%%%%%%%%%%%%%%%%% REFS
%\section*{References}


\begin{thebibliography}{99}

\bibitem{Kivshar1} Y. S. Kivshar and G. P. Agrawal, 
\textit{Optical Solitons: From Fibers to Photonic Crystals} (Academic Press, San Diego, 2003).

\bibitem{Agarwal1} G. P. Agrawal, 
\textit{Applications of Nonlinear Fiber Optics} (Academic Press, San Diego, 2001).

\bibitem{Lakshmanan1} R. Radhakrishnan, M. Lakshmanan, and J. Hietarinta, 
Phys. Rev. E \textbf{56}, 2213 (1997).

\bibitem{Lakshmanan2} T. Kanna and M. Lakshmanan, 
Phys. Rev. Lett. \textbf{86}, 5043 (2001).

\bibitem{Ablowitz1} M. J. Ablowitz, B. Prinari, and A. D. Trubatch, 
Inv. Probl. \textbf{20}, 1217 (2004).

\bibitem{Lakshmanan3} R. Radhakrishnan and M. Lakshmanan, 
J. Phys. A: Math. Gen. \textbf{28}, 2683 (1995).

\bibitem{Kivshar2} A. P. Sheppard and Y. S. Kivshar, 
Phys. Rev. E \textbf{55}, 4773 (1997).

\bibitem{Lakshmanan4} M. Vijayajayanthi, T. Kanna, and M. Lakshmanan,
Phys. Rev. A \textbf{77}, 013820 (2008).

\bibitem{Feng} B. F. Feng,
J. Phys. A: Math. Theor. \textbf{47}, 355203 (2014).

\bibitem{Ohta} Y. Ohta, D. S. Wang, and J. Yang, 
Stud. Appl. Math. \textbf{127}, 345 (2011).

\bibitem{Panos1} P. G. Kevrekidis and D. J. Frantzeskakis, 
Rev. Phys. \textbf{1}, 140 (2016).

\bibitem{Frantzeskakis} D. J. Frantzeskakis, 
J. Phys. A: Math. Theor. \textbf{43}, 213001 (2010).

\bibitem{Kang} J. U. Kang, G. I. Stegeman, J. S. Aitchison, and N. Akhmediev,  
Phys. Rev. Lett., \textbf{76}, 3699 (1996).

\bibitem{Yang99} J. Yang,  
Phys. Rev. E, \textbf{59}, 2393 (1999).

\bibitem{Steiglitz} K. Steiglitz,  
Phys. Rev. E, \textbf{60}, 016608 (2000).

\bibitem{Yang02} J. Yang,  
Phys. Rev. E, \textbf{65}, 036606 (2002).

\bibitem{Zhong} W-P. Zhong, M. Beli, and B. Malomed,  
Phys. Rev. E, \textbf{92}, 053201 (2015).

\bibitem{Zheng2006}
C. Zheng, 
J. Comput. Phys. \textbf{215} 552–565 (2006).
% Exact nonreflecting boundary conditions for one–dimensional cubic nonlinear Schrödinger equations, 

\bibitem{li2011} H. Li, X. Wu, and J. Zhang,
Phys. Rev. E \textbf{84(3)},  036707 (2011).
% Local absorbing boundary conditions for nonlinear wave equation on unbounded domain
%  DOI: 10.1103/PhysRevE.84.036707

\bibitem{Antoine} X. Antoine, Ch. Besse, and S. Descombes, 
SIAM J. Numer. Anal., \textbf{43}, 2272 (2006).

\bibitem{Manakov} S. V. Manakov,
Sov. Phys. JETP  \textbf{38} 248 (1974).

\bibitem{Lakshmanan5} S. Gancsan and M. Lakshmanan,
J. Phys. A: Math. and  Gen. \textbf{20} L1143 (1987).

\bibitem{Lakshmanan6} M. Lakshmanan,
Int. J. Bifurcation Chaos, \textbf{3} 3 (1993).

\bibitem{ismail20081} M.S. Ismail, 
Math. Comp. Simul., \textbf{78} (4), 532 (2008).

\bibitem{Arnold1998} A. Arnold and M. Ehrhardt, 
J. Comput. Phys., \textbf{145(2)}, 611 (1998).

\bibitem{Ehrhardt1999} M. Ehrhardt, 
VLSI Design, \textbf{9(4)}, 325 (1999).

\bibitem{Ehrhardt2001} M. Ehrhardt and A. Arnold, 
Riv. di Math. Univ. di Parma, \textbf{6(4)}, 57 (2001).

\bibitem{Ehrhardt2002} M. Ehrhardt, 
Acta Acustica united with Acustica, \textbf{88}, 711 (2002).

\bibitem{Arnold2003} A. Arnold, M. Ehrhardt, and I. Sofronov, 
Commun. Math. Sci., \textbf{1(3)}, 501 (2003).

\bibitem{Jiang2004} S. Jiang and L. Greengard, 
Comput. Math. Appl., \textbf{47}, 955 (2004).

\bibitem{Antoine2008} X. Antoine, A. Arnold, C. Besse, M. Ehrhardt, and A. Sch\"adle,
Commun. Comput. Phys., \textbf{4(4)}, 729 (2008).

\bibitem{Ehrhardt2008} M. Ehrhardt, 
Appl. Numer. Math. \textbf{58(5)}, 660 (2008).

\bibitem{Matthias2008} A. Zisowsky and M. Ehrhardt, 
Math. and Comput. Modell., \textbf{47}, 1264 (2008).

\bibitem{Sumichrast2009} L. \u{S}umichrast and M. Ehrhardt, 
J. Electr. Engineering, \textbf{60(2)}, 301 (2009).

\bibitem{Antoine2009} X. Antoine, C. Besse, and P. Klein, 
J. Comput. Phys., \textbf{228(2)}, 312 (2009).
% Absorbing boundary conditions for the one-dimensional Schrödinger equation with an exterior repulsive potential

\bibitem{Ehrhardt2010} M. Ehrhardt, 
Numer. Math.: Theor. Meth. Appl., \textbf{3(3)}, 295 (2010).

\bibitem{Klein2011} P. Klein, X. Antoine, C. Besse, and M. Ehrhardt, 
Commun. Comput.  Phys., \textbf{10(5)}, 1280 (2011).

\bibitem{Arnold2012} A. Arnold, M. Ehrhardt, M. Schulte, and I. Sofronov, 
Commun. Math. Sci., \textbf{10(3)}, 889 (2012).

\bibitem{Jambul1} J.R. Yusupov, K.K. Sabirov, M. Ehrhardt, and D.U. Matrasulov,
Phys. Rev. E, \textbf{100}, 032204 (2019).

\bibitem{TBCSGE} K.K. Sabirov, J.R. Yusupov, M. Ehrhardt, and D.U. Matrasulov,
Arxiv:2011.13299 

\bibitem{Jambul} J.R. Yusupov, K.K. Sabirov, M. Ehrhardt, and D.U. Matrasulov,
Phys. Lett. A, \textbf{383}, 2382 (2019).

\bibitem{Jambul02} M.M. Aripov, K.K. Sabirov, and J.R. Yusupov, 
{\it Nanosystems: physics, chemistry, mathematics}, \textbf{10}(5), 501 (2019).

\bibitem{Jambul2} J.R. Yusupov, K.K. Sabirov,  Q.U. Asadov, M. Ehrhardt, and D.U. Matrasulov,
Phys. Rev. E,  \textbf{101(6)}  062208 (2020). 

\bibitem{Exciton} J.R. Yusupov, Kh.Sh. Matyokubov, K.K. Sabirov, and D.U. Matrasulov, 
Chem. Phys., \textbf{537}, 110861 (2020).

\bibitem{ismail2001} M.S. Ismail and T.R. Taha, 
Math. Comp.  Simul., \textbf{56} (6), 547 (2001).

\bibitem{ismail2004} M.S. Ismail and S.Z. Alamri, 
Int. J. Comp. Math., \textbf{81} (3), 333 (2004).

\bibitem{ismail2007} M.S. Ismail and T.R. Taha, 
Math. Comp. Simul., \textbf{74} (4-5), 302 (2007).

\bibitem{ismail2008} M.S. Ismail, 
Math. Comp. Simul., \textbf{196} (1), 273 (2008).

\bibitem{xu2010} Q.-B. Xu and Q.-S. Chang, 
% New numerical methods for the coupled nonlinear Schrödinger equations
Qs. Acta Math. Appl. Sin. Engl. Ser., \textbf{26}, 205 (2010).
% In this paper, three numerical schemes with high accuracy for the coupled Schrödinger equations are studied. The conservative properties of the schemes are obtained and the plane wave solution is analysised. The split step Runge-Kutta scheme is conditionally stable by linearized analyzed. The split step compact scheme and the split step spectral method are unconditionally stable. The trunction error of the schemes are discussed. The fusion of two solitions colliding with different β is shown in the figures. The numerical experments demonstrate that our algorithms are effective and reliable.

\bibitem{frutos1992} J. de Frutos and J.M. Sanz-Serna,
J. Comp. Phys., \textbf{103} (1), 160-168 (1992).

\bibitem{zhou2010}
S. Zhou and X. Cheng,
%"Numerical solution to coupled nonlinear Schrödinger equations on unbounded domains"
Math. Comp. Simul., \textbf{80} (12), 2362 (2010).
% https://doi.org/10.1016/j.matcom.2010.05.019
% Abstract
% The numerical simulation of coupled nonlinear Schrödinger equations on unbounded domains is considered in this paper. By using the operator splitting technique, the original problem is decomposed into linear and nonlinear subproblems in a small time step. The linear subproblem turns out to be two decoupled linear Schrödinger equations on unbounded domains, where artificial boundaries are introduced to truncate the unbounded physical domains into finite ones. Local absorbing boundary conditions are imposed on the artificial boundaries. On the other hand, the coupled nonlinear subproblem is an ODE system, which can be solved exactly. To demonstrate the effectiveness of our method, some comparisons in terms of accuracy and computational cost are made between the PML approach and our method in numerical examples.

\bibitem{chen2017} J. Chen and L.-M. Zhang,
%"Numerical Approximation of Solution for the Coupled Nonlinear Schrödinger Equations"
Acta Math. Appl. Sinica, Engl. Ser., \textbf{33} (2), 435 (2017).
%DOI: 10.1007/s10255-017-0672-3
% (E-mail: cjwan061414@163.com)
% Abstract In this article, a compact finite difference scheme for the coupled nonlinear Schrödinger equations is studied. The scheme is proved to conserve the original conservative properties. Unconditional stability and convergence in maximum norm with order O(τ2 + h4) are also proved by the discrete energy method. Finally, numerical results are provided to verify the theoretical analysis.

\bibitem{hu2020}
Y. Hu, H. Li, and Z. Jiang
% "Efficient semi-implicit compact finite difference scheme for nonlinear Schrödinger equations on unbounded domain",
Appl. Numer. Math., \textbf{153} 319 (2020).

\end{thebibliography}
\end{document}